\documentclass[twocolumn,pre,floats,superscriptaddress]{revtex4-1}

\usepackage{amsmath,amsfonts,amssymb}
\usepackage{graphicx}
\usepackage[]{epsfig}
\usepackage{bm}
\usepackage{dcolumn}
\usepackage{hyperref}

\newcommand{\W}{{\mathcal W}}

\usepackage{color}
\usepackage{comment}
\usepackage[utf8]{inputenc}
\usepackage[english]{babel}

\begin{document}

\title{Numerical test of the replica-symmetric Hamiltonian for
the correlations of the critical state of spin glasses in a field}

\author{L.A.~Fernandez} \affiliation{Departamento de F\'\i{}sica
  Te\'orica, Universidad Complutense, 28040 Madrid, Spain}
\affiliation{Instituto de Biocomputaci\'on y F\'{\i}sica de Sistemas
  Complejos (BIFI), 50018 Zaragoza, Spain}

\author{I.~Gonzalez-Adalid Pemartin} \affiliation{Departamento de
  F\'\i{}sica Te\'orica, Universidad Complutense, 28040 Madrid, Spain}

\author{V.~Martin-Mayor} \affiliation{Departamento de F\'\i{}sica
  Te\'orica, Universidad Complutense, 28040 Madrid, Spain}
\affiliation{Instituto de Biocomputaci\'on y F\'{\i}sica de Sistemas
  Complejos (BIFI), 50018 Zaragoza, Spain}
  
\author{G.~Parisi} \affiliation{Dipartimento di Fisica, Sapienza
  Universit\`a di Roma, P.le A.~Moro 5, 00185 Rome, Italy}
\affiliation{INFN, Sezione di Roma 1, P.le A.~Moro 5, 00185 Rome,
  Italy} \affiliation{CNR-Nanotec, unità di Roma, P.le A.~Moro 5,
  00185 Rome, Italy}

\author{F.~Ricci-Tersenghi} \affiliation{Dipartimento di Fisica,
  Sapienza Universit\`a di Roma, P.le A.~Moro 5, 00185 Rome, Italy}
\affiliation{INFN, Sezione di Roma 1, P.le A.~Moro 5, 00185 Rome,
  Italy} \affiliation{CNR-Nanotec, unità di Roma, P.le A.~Moro 5,
  00185 Rome, Italy}

\author{T.~Rizzo} \affiliation{Institute of Complex Systems (ISC) -
  CNR, Rome unit, P.le A.~Moro 5, 00185 Rome, Italy}

\author{J.J.~Ruiz-Lorenzo} \affiliation{Departamento de F\'{\i}sica,
  Universidad de Extremadura, 06006 Badajoz, Spain}
\affiliation{Instituto de Computaci\'on Cient\'{\i}fica Avanzada
  (ICCAEx), Universidad de Extremadura, 06006 Badajoz, Spain}
\affiliation{Instituto de Biocomputaci\'on y F\'{\i}sica de Sistemas
  Complejos (BIFI), 50018 Zaragoza, Spain}

\author{M.~Veca} \affiliation{Dipartimento di Fisica, Sapienza
  Universit\`a di Roma, P.le A.~Moro 5, 00185 Rome, Italy}

\date{\today}

\begin{abstract}
A growing body of evidence indicates that the sluggish low-temperature
dynamics of glass formers (e.g. supercooled liquids, colloids or spin
glasses) is due to a growing correlation length. Which is the
effective field theory that describes these correlations? The natural
field theory was drastically simplified by Bray and Roberts in
1980. More than forty years later, we confirm the tenets of Bray
and Roberts theory by studying the Ising spin glass in an externally
applied magnetic field, both in four spatial dimensions (data obtained
from the Janus collaboration) and on the Bethe lattice.
\end{abstract}

\maketitle

\section{Introduction}

Spin glasses~\cite{mezard:87,mydosh:93,young:98} in a magnetic field [but
above the de Almeida-Touless (dAT) line~\cite{dealmeida:78}], structural
glasses close to (but above) their mode coupling temperature~\cite{cavagna:09}
or hard spheres above the Gardner transition~\cite{gardner:85}, all display large
correlation lengths and slow relaxations that are typical of a second-order
phase transition. These features are predicted by mean field (MF)
theory~\cite{mezard:87}, and have been identified both in experiments and in
numerical
simulations~\cite{caracciolo:90,huse:91,caracciolo:91,ciria:93,parisi:98b,marinari:98e,marinari:98g,marinari:98h,houdayer:99,marinari:00d,houdayer:00,cruz:03,young:04,leuzzi:08,larson:13,janus:14b,janus:14c,janus:12,leuzzi:09,leuzzi:11,geirhos:18,berthier:05,albert:16,hammond:20,dilucca:20,albert:21}. However,
the very existence of the phase transition has been long
debated~\cite{bray:80,temesvari:02,charbonneau:17,angelini2022unexpected}. Indeed, it has
been frequently suggested that these \emph{critical} features might be
connected to a crossover, rather than to a true
phase-transition~\cite{bray:87,bray:87b,fisher:85,fisher:86,fisher:88,fisher:88b,fischer:93}:
the corrections to MF theory would destroy the transition, or (in some cases)
move it to zero temperature.  In this paper we do not claim against, nor in
favor, of the presence of a transition. Instead, our aim is understanding in
detail the properties of the correlations in the region where the
susceptibilities are large (e.g. $10^3$ times their natural value).

Let us consider the framework of spin glasses in a
magnetic field. The theory is complex~\cite{parisi:13}.  Three different
two-point correlators (and their associated susceptibilities) become
critical. We also have eigth non-linear susceptibilities associated to the the
eight three-point correlators (there are eight different coupling constants)
\footnote{We call susceptibilities (non-linear susceptibilities), to the
  two-point (three-point) correlators in Fourier space, at zero momentum.}.
However, in an expansion around MF, one finds a linear transformation such
that only one of the three susceptibilities is divergent at the critical
temperature $T_{\text{c}}$. Similarly, the divergence at $T_\mathrm{c}$ is
more violent for two of the non-linear susceptibilities: at first order in
perturbation theory, they scale as $1/(T-T_\mathrm{c})^3$ , while two
non-linear susceptibilities diverge as $1/(T-T_\mathrm{c})^2$, another one as
$1/(T-T_\mathrm{c})$ and the remaining three are finite at $T_{\text{c}}$.  As
expected, only the couplings that correspond to the most divergent non-linear
susceptibilities are relevant near the transition. The linear transformations
that {\it diagonalize} the singularity structure are well known, and they have
a physical meaning. Corrections to MF could completely destroy this
divergences structure (or they may just modify the values of the critical
exponents). A systematic investigation of the correctness of the above
picture has never been attempted using numerical simulations.  This paper
fills the lacuna in the particular case of spin glasses. We show that these
qualitative predictions are satisfied in the region of large
susceptibilities. It is quite possible that the same situation is present in
other contexts, beyond spin glasses.

\section{Summary of the theoretical framework}

The standard tool to understand the fate of a transition in finite spatial
dimension $D$ is the Wilsonian Renormalization Group
(RG)~\cite{wilson:74}. Unfortunately, the standard perturbative construction
fails in these models: the most relevant corrections to MF theory are due to
the presence of cubic terms in the effective Landau-Ginsburg theory (LGT), see
Eq.~(\ref{T3}) in Appendix~\ref{appA}, and two coupling ($\widetilde{w}_1$ and
$\widetilde{w}_2$) are known to be relevant for $D\lesssim 6$. In fact, in
spin glasses and also in models with the same LGT, the construction of the
$D=6-\epsilon$ expansion fails because no fixed point is present in the
weak-coupling region~\cite{bray:80}. The action of the RG brings the
corrections to the Gaussian behavior in the region where the effective
couplings are large. The fate of the parameter
$\lambda_r\equiv w_{2,r}/w_{1,r}$
($w_{i,r}$, $i=1,2$, are the renormalized couplings, see
below) is of particular interest. Indeed, $\lambda_r$ plays a
crucial role in the mode coupling theory where it must be
$0\leq \lambda_r\leq 1$. Moreover, as discovered by Gross et
al.~\cite{gross:85}, and recently stressed by H\"oller and
Read~\cite{holler:20}, having $\lambda_r>1$ would imply a peculiar first-order
like transition, like the calorimetric transition of glasses (see
e.g.~\cite{cavagna:09}).

Unfortunately, in spite of the relevance of the renormalized
parameters~\footnote{As usual the renormalized couplings are defined in terms
  of the renormalized correlation functions at zero momentum.}, they have not
been obtained in simulations, partly because of the complexity of the
computation. Here we show that such a computation is feasible: we present
results for spin glasses in a magnetic field, both in the
Bethe lattice and in the $D=4$ hypercubic lattice. Our model choice is based
on its relative simplicity, but our techniques can be straightforwardly
extended to more complex models. The Bethe lattice computation is a test of
the viability of the approach and of the formulae used. Indeed, corrections
to MF disappear in an infinite Bethe lattice and the value of $\lambda_r$,
which is unaffected by fluctuations (i.e. loop corrections), is analytically
known. On the other hand, the $D=4$ Edwards-Anderson (EA) mode may be well
thermalized in the region of very large susceptibilities and we have some
estimates of the position of the extrapolated dAT transition~\cite{janus:12}.
Our results are suggestive of the presence of a fixed-point value
$\lambda_r\approx 0.5$, and clearly exclude a value of $\lambda_r$ greater
than 1.

Let us summarize the theoretical understanding for spin glasses in a magnetic
field $h$. The effective action can be written using the replica formalism
(we recall in  Appendix \ref{appA} the main results, that are well described in
the literature). We aim to express all our results in terms of correlation
functions than can be computed in a numerical simulation. Let us start from
the two points correlation functions. As usual in disordered systems, we need
to distinguish between the thermal average, $\langle (\cdots) \rangle$, and
the average over disorder, $\overline {(\cdots)}$.  For a system of linear
size $L$, with $N=L^D$ spins $S_i=\pm 1$, we have three relevant
susceptibilities:
\begin{eqnarray}\chi_1& \equiv & \frac{1}{N} \sum_{ij} \overline{\langle S_i \,
  S_j \rangle^2} -q^2\,,\\ \chi_2& \equiv & \frac{1}{N} \sum_{ij}
\overline{\langle S_i \, S_j \rangle^ \langle S_i\rangle \langle
  S_j\rangle }-q^2\,,\\ \chi_3& \equiv & \frac{1}{N} \sum_{ij}
\overline{\langle S_i\rangle^2 \langle S_j\rangle^2}-q^2\,, \end{eqnarray}
where $q\equiv \overline{\langle S_i\rangle^2}$ is the average
overlap.  If we expand around the MF solution we find at all
orders of the perturbation theory that the so-called replicon
susceptibility is divergent near the transition:
\begin{equation} \chi_R \equiv
\chi_{SG} \equiv \frac{1}{N} \sum_{ij} \overline{\langle S_i \, S_j
  \rangle_c^2} =\chi_1-2 \chi_2+\chi_3\,,\label{eq:chiR-def}
\end{equation}
where by $\langle (\cdots) \rangle_c$ we denote the connected correlation
function (e.g.
$\langle S_i \, S_j \rangle_c=\langle S_i\,S_j\rangle -\langle
S_i\rangle\langle S_j\rangle$, see for instance ~\cite{parisi:88}).  For later
use we introduce the longitudinal  and anomalous
susceptibilities, $\chi_L$ and $\chi_A$. The two are degenerated in presence of a magnetic field,
\begin{equation}\label{eq:chiL-def}
\chi_L=\chi_A=\chi_1-4 \chi_2 + 3 \chi_3\,.
\end{equation}

If we consider Gaussian-distributed random magnetic fields, $\chi_L$ is
proportional to the staggered magnetic
susceptibility (see Appendix \ref{appC} for a detailed discussion). Then, the physically motivated assumption
that the magnetic susceptibility is not critical implies that $\chi_L$ is not
critical either. Only the average of the (squared) connected-correlator
becomes critical. This is in sharp contrast, with the $h=0$ case where
$\chi_2=\chi_3=0$ and $\chi_A=\chi_L=\chi_R$.  We expect a crossover region
for small $L$ and $h$, where $\chi_L$ and $\chi_A$ seem critical (because
$\chi_L$ and $\chi_A$ \emph{are} critical at the $h=0$ transition).

The renormalized coupling $w_{1,r}$ and $w_{2,r}$ are defined in terms of the exact vertices $w_1$ and $w_2$, {\it i.e.} the coefficients of the Gibbs Free energy. The exact vertices can be expressed as  $w_i = \omega_i/\chi_R^3$ ($i=1,2$) in terms of 
connected-correlations at zero external momentum~\cite{parisi:13},
  \begin{eqnarray} \omega_1 & \equiv &
 \frac{1}{N} \sum_{ijk} \overline{\langle S_i \, S_j \rangle_c\,
   \langle S_j \, S_k \rangle_c \, \langle S_k \, S_i \rangle_c}
 \,,\label{eq:omega1def}\\
 \omega_2 & \equiv & \frac{1}{2\,N} \sum_{ijk} \overline{\langle
   S_i \, S_j\, S_k \rangle^2_c}\,,\label{eq:omega2def} \end{eqnarray}
   The coupling constants $w_1$, $w_2$ diverge at the transition while the renormalized coupling constants
remain finite. They are obtained renormalizing the lengths and the overlap fields leading to:
\begin{equation}
  w_{1,r} = \frac{\omega_1}{\chi_{R}^{3/2} \, \xi_2^{D/2}}\, \ , \
  w_{2,r} = \frac{\omega_2}{\chi_{R}^{3/2} \, \xi_2^{D/2}}\, \,,
\label{wren}
\end{equation}
where $\xi_2$ is the second-moment correlation length. It follows that
\begin{equation}
\lambda_r=\frac{w_{1,r}}{w_{2,r}}=\frac{\omega_1}{\omega_2}\,.
\label{eq:def-lambda}
\end{equation}
Note that $\lambda_r=w_1/w_2$: hence,
$\lambda$ does not renormalize and we will drop thereafter the
sub-index $r$.

Finite-volume corrections are very strong so we do not consider here
the computation of the renormalized couplings $w_{i,r}$. However,
we can introduce the dimensionless quantities
\begin{equation}\label{eq:Lambda-def}
  \Lambda_1=\frac{\omega_1}{\chi_R^{3/2} L^{D/2}} \,,\qquad
  \Lambda_2=\frac{\omega_2}{\chi_R^{3/2} L^{D/2}} \,.
\end{equation}
that should scale with $L$ as Binder's cumulant~\cite{binder:81}. Notice that,
at the critical point, $\Lambda_i \propto w_{i,r}$.

Before discussing our numerical findings for $\lambda$,
it is important to stress
that there are nonequivalent ways of taking the relevant limit for
$\lambda(L,T)$ in the onset of a second order phase transition at $T_c$:
\begin{equation}\label{eq:limits}
  \lambda^*=\lim_{L\rightarrow \infty} \lim_{T \rightarrow T_c} \lambda(L,T)\,,\quad
  \lambda(T_c^+)= \lim_{T \rightarrow T_c^+} \lim_{L\rightarrow
  \infty} \lambda(L,T)\,.
\end{equation}
The fact that $\lambda^*\neq \lambda(T_c^+)$ is hardly
surprising~\cite{salas:00}. Similarly, the corresponding limits for
renormalized coupling $w_{1,r}$ and $w_{2,r}$ do not
commute.  $\lambda(T_c^+)$ is in general more difficult to estimate than
$\lambda^*$, but the former could be more desirable given that the RG
$\beta$-functions(see, e.g.,~\cite{parisi:88,zinn-justin:05,amit:05}) are typically expressed in terms of the thermodynamic quantities in
analytical computations.

\section{Numerical simulations results}

In a simulation, the above quantities are computed from real \emph{replicas} (i.e.
systems that evolve independently under the same coupling constants). It is
well known that one needs {\it
  two} real replicas to compute $q$, {\it four} replicas for the three
susceptibilities, and {\it six} replicas for the $\omega_{i=1,2}$ in
Eqs.~(\ref{eq:omega1def}\,,\ref{eq:omega2def}). In spite of this and {\it only
  at the critical point}, it is possible to compute both $\omega_i$ using only
three and four replicas.  We shall denote the estimate
obtained with $R$ replicas by $\omega_i^{(R=3,4)}$. 
Away from the critical point, one has for the differences
$\omega_i-\omega_i^{(R)}=O(|T-T_c|^{\rho(R)})$,
where $\rho(R)$ is a suitable exponent (see Appendix \ref{appB} for a more complete discussion).

\paragraph*{Numerical results in the Bethe lattice.}

To study the behavior of the three- and four-replicas estimators in a
controlled setting, we have simulated an Ising spin glass in a magnetic field
on a Bethe lattice (random regular graph with fixed-degree 4). In this case,
there is little doubt that a true dAT transition is present. Furthermore, the
divergence of the susceptibilities (both linear and non-linear) closely matches 
our description above.

\begin{figure}[t]
\begin{center}
\includegraphics[width=\columnwidth]{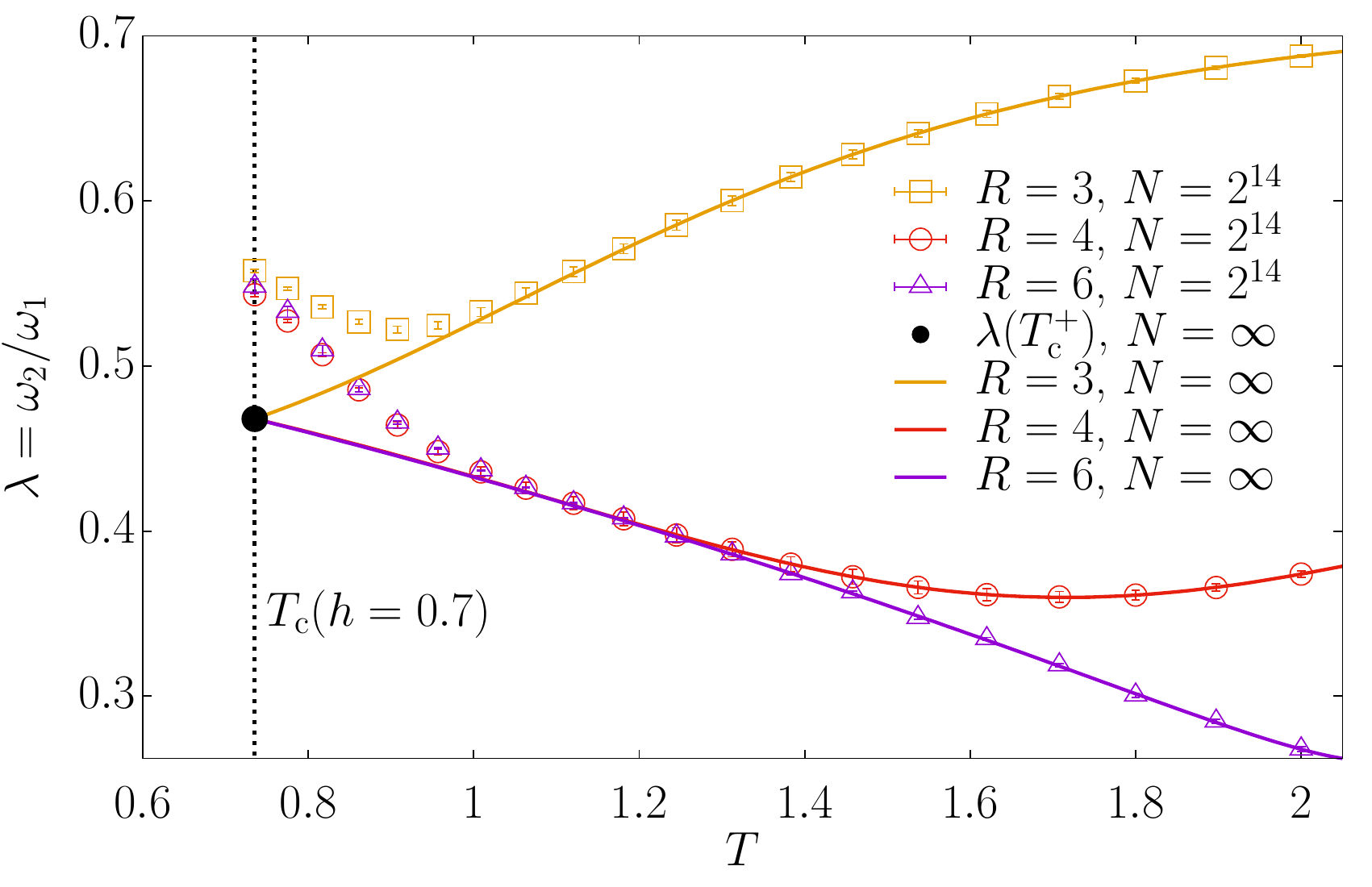}
\caption{Temperature dependence of the ratio of renormalized
  couplings $\lambda$, see Eq.~(\ref{eq:def-lambda}),
  computed with a magnetic field $h=0.7$ on a Bethe lattice.
  The critical temperature is marked with a vertical line. We
  plot the data obtained with the three-, four- and six-replicas
  estimators. The black dot reports the value of $\lambda(T_c^+)\simeq 0.47$, see
  Eq.~(\ref{eq:limits}), that has been computed analytically in \cite{parisi:14}. All three
  estimators take the same value $\lambda^*\simeq 0.55$ at the critical temperature.
  The continuous lines, marked with
  $N=\infty$, are the extrapolations of the data considering scaling
  corrections~\cite{veca:20} and are compatible with the analytical
  computation $\lambda(T_c^+)$.}
\label{Fig:Bethe}
\end{center}
\end{figure}

In Fig.~\ref{Fig:Bethe} we plot the parameter $\lambda$ for the Bethe lattice,
as obtained from the exact expression together with the three- and four-replica
estimators $\lambda^{(3)} \equiv \omega_2^{(3)}/\omega_1^{(3)}$ and
$\lambda^{(4)}\equiv \omega_2^{(4)}/\omega_1^{(4)}$.  In this case $T_c$ and
$\lambda(T_c^+)$ are known analytically \cite{parisi:14} and we see that the
estimators extrapolate to the correct value at the critical temperature,
although close to the critical point there are finite size corrections. Note
as well that the finite-size corrections of the the true $\lambda$ (i.e. the
six-replica estimator) and of the four-replica estimator coincide in the
critical region. The same effect is expected for the three-replica estimator
but it is masked by pre-asymptotic effects at the sizes considered.  At any
rate, we find that the deviations are consistent with the predicted MF values
$\omega_i-\omega_i^{(3)} = O(|T-T_c|)$ and
$\omega_i-\omega_i^{(4)} = O(|T-T_c|^{3})$ \cite{veca:20}.

\paragraph*{Numerical results in four dimensions.}

The discussion of the three- and four-replica estimators is of great
practical and theoretical importance in this case.

The theoretical importance relies on the fact that, at variance with the Bethe
lattice case, one cannot take for granted that the transition is described by
the theory outlined above. For instance, we could have a continuous transition
described by a different theory and therefore the three- and four-replica
estimators would yield conflicting results, thus indicating a wrong choice for
the starting field-theory. Furthermore, due to the lack of a perturbative RG
fixed-point below six dimensions, one could even question the very existence
of such a theory for $D<6$.  Thus the fact that the three- and four-replica
expressions yield consistent estimates provides a non trivial indication that
the region of large susceptibilities is actually described by the
Replica-Symmetric field theory of Bray and Roberts~\cite{bray:80}.
\begin{figure}[t]
\begin{center}
\includegraphics[width=\columnwidth]{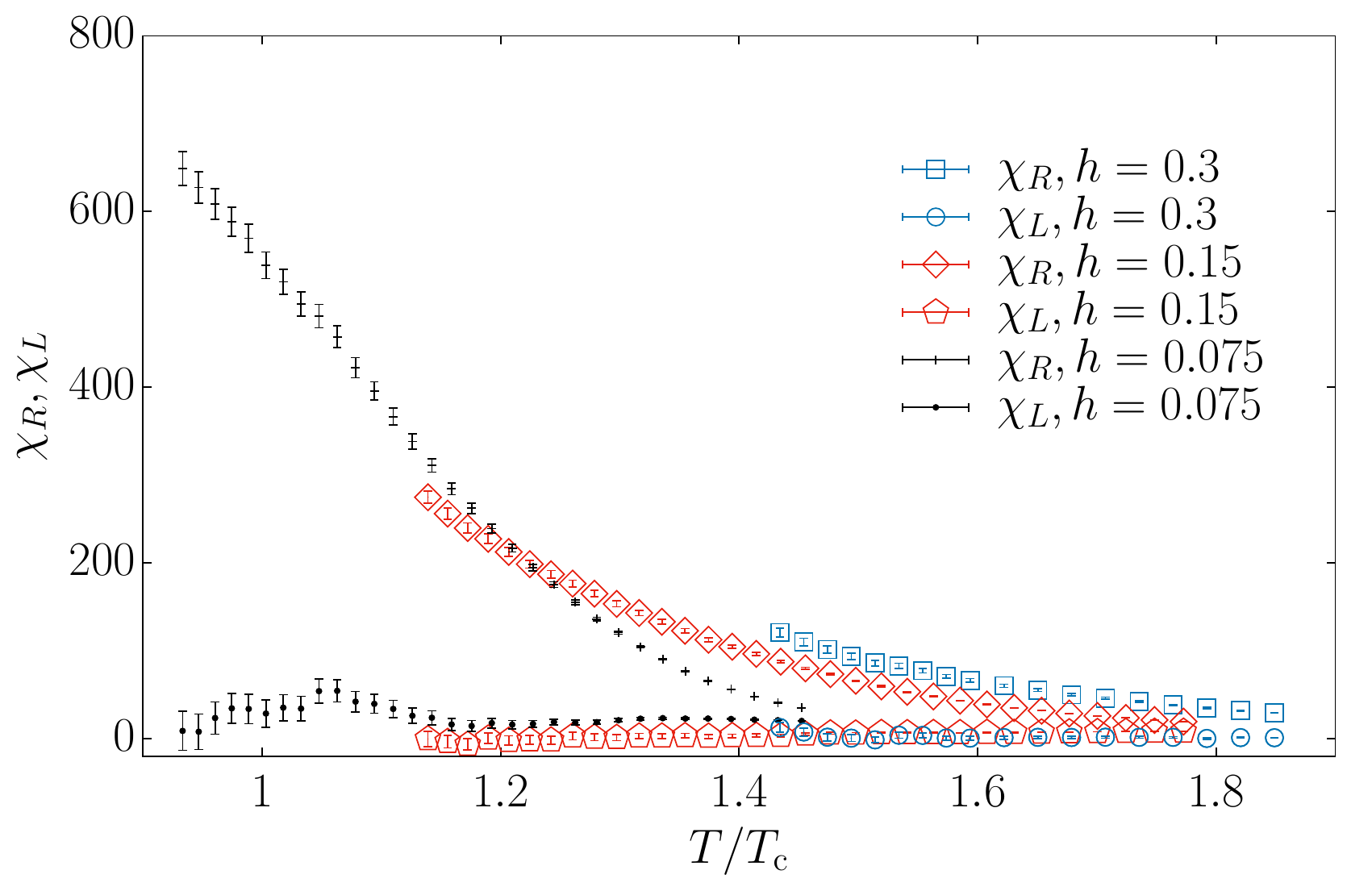}
\caption{Replicon ($\chi_R$) and longitudinal ($\chi_L$) susceptibilities,
  Eqs.~(\ref{eq:chiR-def},\ref{eq:chiL-def}) vs. temperature, as computed for
  the $D=4$ Edwards-Anderson model at magnetic fields $h=0.075, 0.15$ and
  $0.3$ (for each $h$, the temperature is plotted rescaled by the corresponding best estimate for $T_c$ \cite{janus:12}). To avoid cluttering the plot, we only
  show data for our largest system, $L=16$.}
\label{Fig:chi}
\end{center}
\end{figure}

The practical importance of the three- and four-replica estimators
lies in that, in the present study, we have re-analyzed equilibrium
configurations obtained by the Janus Collaboration~\cite{janus:12}
using the Janus-I supercomputer~\cite{janus:08} where the
four-dimensional Ising spin glass in presence of a constant magnetic
field was simulated (see Appendix \ref{appD}).  Those
equilibrium configurations were obtained only for four real
replicas. Therefore, $\lambda$ can be computed only through the three-
and four-replica estimators (although the computation will not be
exact away from the dAT line).

In~\cite{janus:12} the critical temperatures and the critical exponents were
estimated for three different magnetic fields ($h=0.075,0.15$ and $0.3$), by
looking only to one of the two-point correlators, namely the replicon. We
study the same magnetic fields considered in \cite{janus:12}, for temperatures
near (but above) their estimated critical temperatures.

We start by studying in Fig.~\ref{Fig:chi} the replicon and longitudinal susceptibilities, recall Eqs.~(\ref{eq:chiR-def}) and (\ref{eq:chiL-def}).
We clearly see that $\chi_R$ increases and becomes very large as the temperature is lowered, while $\chi_L$ saturates at a much smaller plateau value~\footnote{The plateaux value of $\chi_L$ in Fig.~\ref{Fig:chi} approximately scale as $h^{-x}$, with $x$ between 2 and 3. In MF, $\chi_L$
near $T_\mathrm{c}$ is proportional to
$h^{-2/3}$~\cite{dedominicis:12}.}.
We conclude, in agreement with our MF-based expectations and with previous dynamic investigations in $D=3$~\cite{janus:14b}, that correlations extend to much larger distances for the replicon mode than for the longitudinal one, thus excluding the possibility that the critical behavior in $\chi_R$ is due to the $h=0$ fixed point.

\begin{figure}[t]
\begin{center}
\includegraphics[width=\columnwidth]{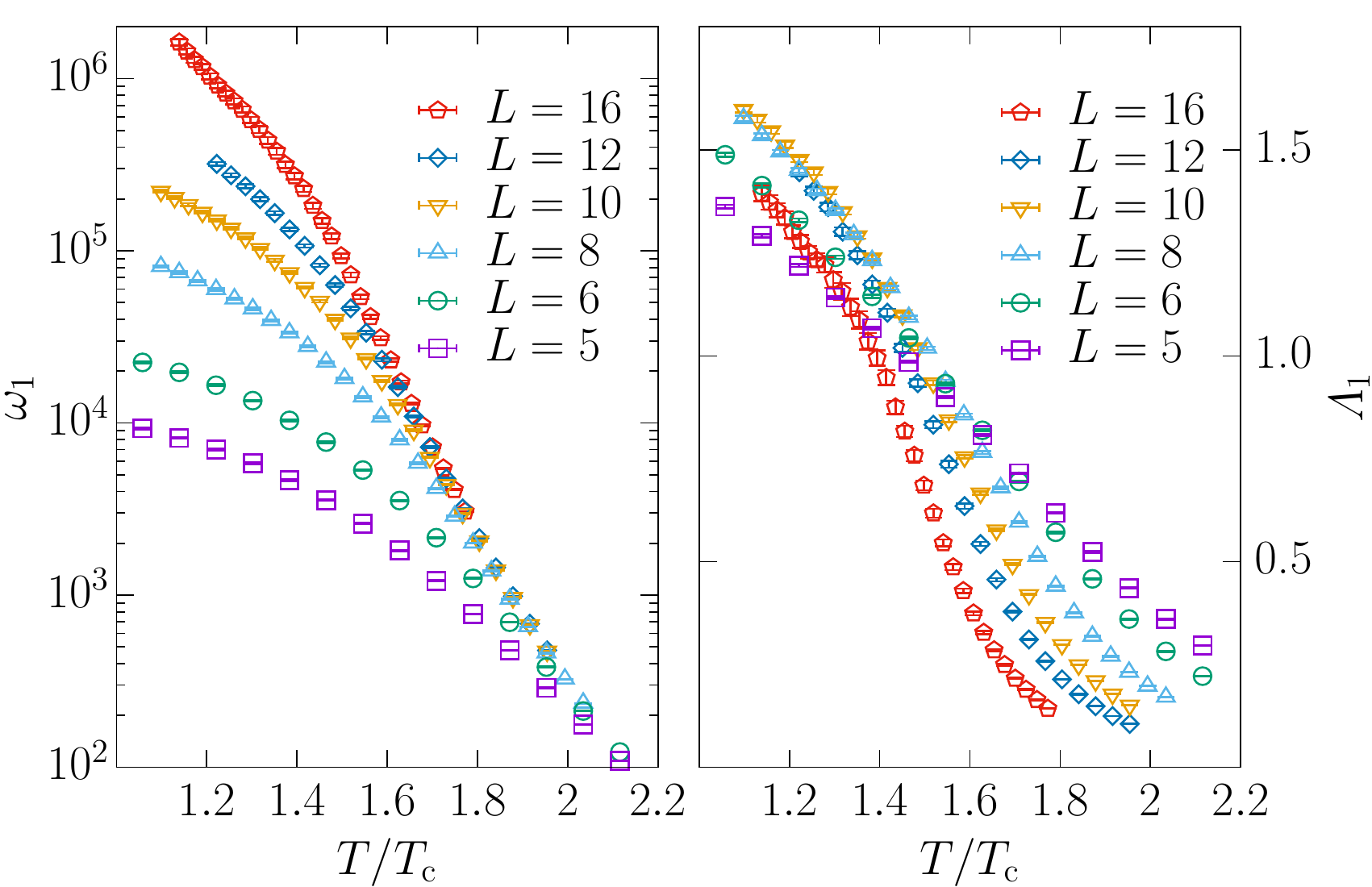}
\caption{$D=4$ Edwards-Anderson model with magnetic field    $h=0.15$. {\bf Left:} Four-replica estimate, $\omega_1^{(4)}$, for the non-linear susceptibility in Eq.~\eqref{eq:omega1def} vs. temperature.
  {\bf Right:} Dimensionless quantity $\Lambda_1$, recall Eq.~\eqref{eq:Lambda-def}, vs. temperature. In both panels, $T$ is in units of the estimated critical temperature $T_c$~\cite{janus:12}.}
\label{Fig:omega1}
\end{center}
\end{figure}

We have considered also the non-linear susceptibilities, the most divergent ones being $\omega_1$ and $\omega_2$, see Eqs.~(\ref{eq:omega1def},\ref{eq:omega2def}).
We find that $\omega_1$ grows significantly upon decreasing $T$ and (at a fixed, low $T$) upon increasing $L$, see Fig.~\ref{Fig:omega1} (left).
The suggested divergence in $\omega_1$ makes it advisable to consider the dimensionless $\Lambda_1(L,T)$, see Eq.~\eqref{eq:Lambda-def}.
At a critical point, the curves of $\Lambda_1$ as function of $T$, computed for different sizes $L$, should cross or merge at $T_\mathrm{c}$. Our data for $L=10,12$ and $16$ in Fig.~\ref{Fig:omega1} (right) do not clearly cross nor merge, making it difficult to compute $T_{\mathrm{c}}$ from these data (indeed, the authors of Ref.~\cite{janus:12} could locate $T_{\mathrm{c}}$ only by considering quantities at non-zero external momentum).
The crucial point, however, is the absence of any evidence in Fig.~\ref{Fig:omega1} (right) for a runaway trajectory where $\Lambda_1$ becomes bigger and bigger upon increasing $L$.
This observation makes unlikely the scenario with a first order transition~\cite{holler:20}.

\begin{figure*}[t]
\begin{center}
\includegraphics[width=\textwidth]{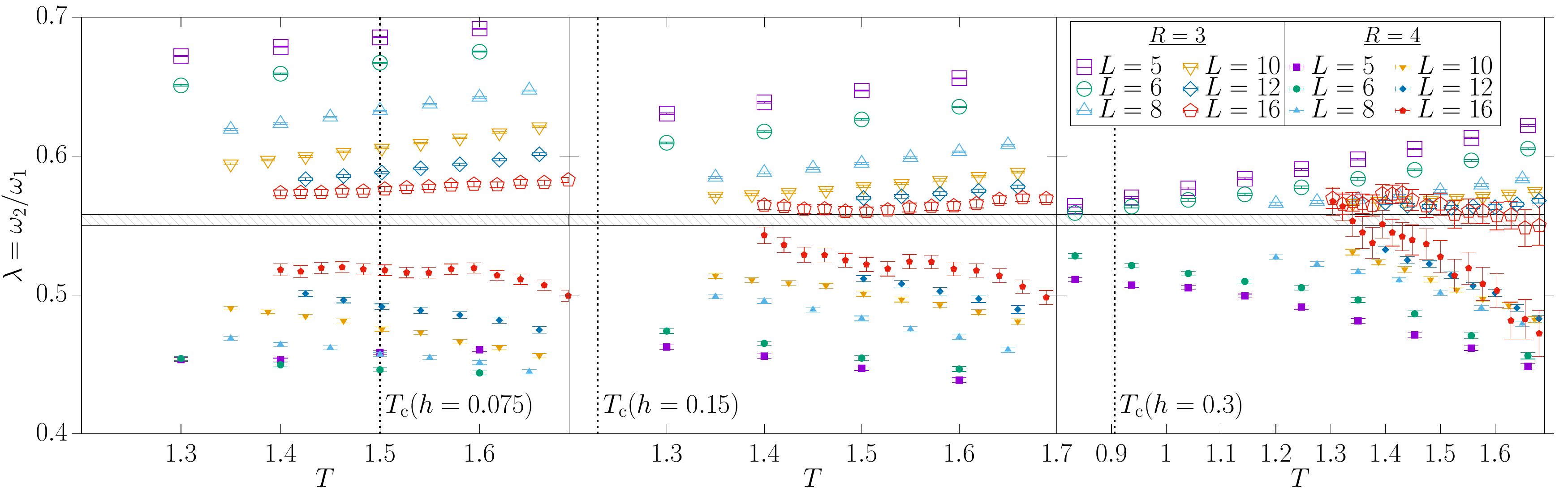}
\caption{Three- and four-replicas estimators for $\lambda$ as a function
  of the temperature in the $D=4$ Ising spin glass (the value
  of the magnetic field is indicated above each panel). Vertical lines
  report the three critical temperatures taken from \cite{janus:12}.
  The band around $\lambda^*\simeq 0.55$ is our best $L\to\infty$
  extrapolation, 
  assuming three- and four-replicas estimators converge to a common value
  for all the three simulated values of the magnetic field (the width of
  the band represents the uncertainty in our extrapolation for $h=0.075$).}
\label{Fig:lambda4D}
\end{center}
\end{figure*}

Once we know that $\omega_{1,2}$ behave as expected, we can consider their
ratio $\lambda$, which is the main quantity of interest.
Fig.~\ref{Fig:lambda4D} shows the three- and four-replica estimators for
magnetic fields $h=0.075$, $0.15$ and $0.30$. At variance with our findings
for the Bethe lattice (where the difference between $\lambda(T_c^+)$ and
$\lambda^*$ is very clear, recall Eq.~\eqref{eq:limits} and
Fig.~\ref{Fig:Bethe}), our data for the $4D$ case shown in
Fig.~\ref{Fig:lambda4D} do not manifest large finite size effects approaching
the critical point: data barely depend on temperature for $T<T_c(h=0)$, thus
suggesting $\lambda^*$ and $\lambda(T_c^+)$ should be very close. The only
visible finite size effect in $4D$ data is a monotonic in $L$ decrease for
$R=3$ and increase for $R=4$, that actually helps in bracketing $\lambda^*$
between the values measured on the largest lattice $L=16$. Indeed, our data
are consistent with a universal value $\lambda^* \approx 0.55$ at the critical
temperature. We remark as well that both the $R=3$ and the $R=4$ estimates
verify $\lambda(L,T)<1$. Hence, we conclude $\lambda(T_c^+)<1$ in $4D$ spin
glasses in a field, which is the main result of this paper.

\section{Conclusions}

Irrespectively of the ongoing debate about whether the glass transition is a
true phase transition or a crossover, it is undeniable that glass formers
display slow dynamics and large correlations. When the lengthscale for
fluctuations becomes large, the natural tool to study the problem is a Field
Theory. Unfortunately, symmetry considerations do not constraint much the
Hamiltonian: in the particular case of spin glasses in a magnetic field, we
end with a extremely complex theory containing eight different
coupling-constants. Bray and Roberts~\cite{bray:80} drastically simplified the
theory. Their so-called replica symmetric Hamiltonian has been the basis for
many analysis. In spite of this, up to now it was not possible to test in a
non-trivial problem the basic hypothesis underlying the theory.  We have
overcome this challenge thanks to two crucial ingredients: (i) a detailed
scaling description for the many linear and non-linear susceptibilities in the
problem~\cite{parisi:13}, and (ii) a re-analysis of the equilibrated
configurations obtained with the Janus I supercomputer~\cite{janus:12}.  We
have found that the crucial scaling relations are fullfilled beyond Mean Field
approximation, close (but above) the de Almeida-Thouless line. Furthermore, it
is quite probable that our approach will be relevant for the study of other
physical systems as well (e.g. glass-forming liquids).  Besides, our results
for the renormalized coupling $\lambda$ seem to exclude the suggested scenario
of a first-order transition~\cite{holler:20}.

\begin{acknowledgments}
The authors wish to thank the Janus Collaboration for allowing us to
analyze their data.  We would like also to thank E.~Marinari for
interesting discussions. The analysis of the Janus configurations was
performed at ICCAEx supercomputer center in Badajoz, we thank its
staff for their assistance.

This work was supported by the European Research Council under the
European Unions Horizon 2020 research and innovation programme (grant
No. 694925, G. Parisi), by Ministerio de Econom\'{\i}a y
Competitividad, Agencia Estatal de Investigaci\'on, and Fondo Europeo
de Desarrollo Regional (FEDER) (Spain and European Union) through
grants No.~PID2020-112936GB-I00 and No.~PGC2018-094684-B-C21, and by
Junta de Extremadura (Spain) through grants No.~GRU18079 and
No.~IB20079 (partially funded by FEDER).  IGAP was supported by MCIU
(Spain) through FPU grant No.~FPU18/02665.
\end{acknowledgments}

\appendix

\section{The replica-symmetric field theory.}
\label{appA}

Standard arguments~\cite{bray:80,temesvari:02} tell us that
the $D$-dimensional Ising spin glass in presence of a magnetic field
is described at criticality by the following RS
Hamiltonian for the replicated overlap $\phi_{ab}(x)$
($\phi_{aa}(x)=0$):
\begin{eqnarray} {\mathcal H} & = & \frac12\int d^Dx
\left[ m_1 \sum_{ab}\phi_{ab}^2+\frac12 \sum_{ab} (\nabla
  \phi_{ab})^2+ \right.  \nonumber \\ &+&m_2\sum_{abc}
  \phi_{ab}\phi_{ac}+ m_3\sum_{abcd}\phi_{ab}\phi_{cd} + \nonumber
  \\ & - &\left. \frac16 \widetilde{w}_1
  \sum_{abc}\phi_{ab}\phi_{bc}\phi_{ca}- \frac16 \widetilde{w}_2
  \sum_{ab}\phi_{ab}^3\right]\,.
\label{T3}
\end{eqnarray}
Note that the cubic coupling in the Hamiltonian are written as $\widetilde{w}_1$, $\widetilde{w}_2$. In general they are different from the corresponding coefficients $w_1$,$w_2$ of the Gibbs Free energy discussed in Ref.~\cite{parisi:13} (the vertices in field theoretical language). The Gibbs free energy as usual is the Legendre transform of the free energy, the corresponding coefficients of the free energy are $\omega_1$, $\omega_2$ introduced before.
The coefficients $\widetilde{w}_i$ and $w_i$ are respectively bare and
dressed couplings and they coincide only at the level of the tree
approximation in field theory, in general they are different.

At the MF level (where $\widetilde{w}_i=w_i$), $m_1$ vanishes linearly on the dAT
line and, in the SG phase, the solution displays
Replica-Symmetry-Breaking (RSB) with a breaking point at a value equal
to $w_2/w_1$ \cite{gross:85,rizzo:13}: it
follows that $\lambda \equiv w_2/w_1$ must be
smaller than one for consistency. It should be also noted that the
parameter $\lambda$ controls the MF values of equilibrium and
off-equilibrium dynamical exponents in a variety of
contexts~\cite{caltagirone:12,parisi:13,caltagirone:13}.

The idea of H\"oller and Read~\cite{holler:20} (that started
from~\cite{moore:18}), is to apply the RG to the above replicated
Hamiltonian until the mass term $m_1$ (which is initially small
because we start close to the dAT line) becomes equal to one, then the
RG flow is stopped and the new Hamiltonian is analyzed at the MF
level. Note that they actually follow Bray and Roberts \cite{bray:80}
and project on the replicon subspace effectively sending the
longitudinal and anomalous masses to infinity. To obtain subcritical
behavior one must keep the massive modes finite, see
ref. \cite{dedominicis:02,dedominicis:12} for a thorough comparative
discussion of the two approaches.  H\"oller and Read suggest that
below the upper critical dimension, $\lambda$ becomes larger than one
under the RG flow on the whole dAT line and therefore the transition
becomes first-order. One should note that treating a Wilson
Hamiltonian at the MF level is always an approximation, although it
may be accurate close to the upper critical dimension. Essentially,
one is approximating the true Gibbs free energy with the Wilson's
Hamiltonian, {\it i.e.}  fluctuations are neglected.  While the
coefficients of the Wilson's Hamiltonian are bare parameters that
cannot be measured, the coefficients of the Gibbs free energy
(proportional of the renormalized couplings) can be expressed in terms
of physical observables and thus are directly accessible to
measurements~\cite{parisi:88,zinn-justin:05,amit:05}.

The renormalized couplings $w_{1,r}$ and
$w_{2,r}$ have finite and model-dependent values except at
the critical temperature where, if scaling holds, they have {\it
  finite} universal values $w_{1,r}^*$ and
$w_{2,r}^*$.  The spin-glass susceptibility and
correlation length diverge as $\chi_R \propto |T-T_c|^{-\gamma}$ and
$\xi_2 \propto |T-T_c|^{- \nu}$ respectively, and consistently
$\omega_1$ and $\omega_2$ diverge as:
\begin{equation}
\label{scaling_omega3}
\omega_{1,2} \propto |T-T_c|^{-\gamma_3}\;, \quad \gamma_3= 3 \nu - \frac{3}{2} \nu \,
\eta+ \frac{\nu D}{2}\;.
\end{equation}
Notice that renormalized couplings constants $w_{1,r}^*$ and $w_{2,r}^*$ are
universal quantities at criticality and play a key role in
computations of critical exponents
\cite{parisi:88,zinn-justin:05,amit:05}, being the zeroes of the
$\beta$-functions.

Note that Eqs.~(8) in the main text follow from Eqs.~(91) in
Ref.~\cite{parisi:13} noticing that when the RG flow is stopped the overlaps
are effectively rescaled by a factor $\chi_R^{1/2}$ and the length is rescaled
by a factor $\xi_2$ since the coefficient of the term $(\nabla\phi_{ab})^2$ is
fixed to one in the RG flow.

\section{Computing $\omega_1$ and $\omega_2$ using three, four and six replicas.}
\label{appB}
In order to compute $\omega_1$ and $\omega_2$ we need to evaluate
numerical quantities like
\begin{equation}
  m^2_i\equiv\overline{\langle\sigma_i\rangle^2} \,,
  \quad m^4_i\equiv\overline{\langle\sigma_i\rangle^4}\,,  \quad
  m^6_i\equiv\overline{\langle\sigma_i\rangle^6}\,.
\end{equation}
The standard approach consists in introducing $K$ independent replicas of the system sharing the same disorder ($\sigma^{(i)}$, $i=1,\dots,K$) obtaining
\begin{equation}\nonumber
  m^2_i=\overline{\langle\sigma^{(1)}_i\sigma^{(2)}_i\rangle} \,,\,
  m_i^4=\overline{\langle\sigma^{(1)}_i\sigma^{(2)}_i\sigma^{(3)}_i\sigma^{(4)}_i\rangle} \,,\,\\ 
\end{equation}
\begin{equation}
  m_i^6=\overline{\langle\sigma^{(1)}_i\sigma^{(2)}_i\sigma^{(3)}_i\sigma^{(4)}_i\sigma^{(5)}_i\sigma^{(6)}_i\rangle}\,.
\end{equation}

Both non-linear susceptibilities, $\omega_1$ and $\omega_2$, are
suitable for numerical evaluation once expressed as \cite{parisi:13}
\begin{eqnarray*}
\omega_1 & = & \W_1 -3 \W_5+3 \W_7-\W_8\,,\\
\omega_2 & = & \frac12 \W_2 - 3 \W_3 + \frac32 \W_4 + 3 \W_5 + 2 \W_6
-6 \W_7+2 \W_8 \,,
\end{eqnarray*}
and
\begin{eqnarray*}
\W_1 & \equiv & N^2 \overline{\langle \delta Q_{12} \delta Q_{23} \delta Q_{31} \rangle}\,,\\
\W_2 & \equiv & N^2 \overline{\langle \delta Q_{12}^3 \rangle}\,,\\
\W_3 & \equiv & N^2 \overline{\langle \delta Q_{12}^2 \delta Q_{13} \rangle}\,,\\
\W_4 & \equiv & N^2 \overline{\langle \delta Q_{12}^2 \delta Q_{34} \rangle}\,,\\
\W_5 & \equiv & N^2 \overline{\langle \delta Q_{12} \delta Q_{13} \delta Q_{24} \rangle}\,,\\
\W_6 & \equiv & N^2 \overline{\langle \delta Q_{12} \delta Q_{13} \delta Q_{14} \rangle}\,,\\
\W_7 & \equiv & N^2 \overline{\langle \delta Q_{12} \delta Q_{13} \delta Q_{45} \rangle}\,,\\
\W_8 & \equiv & N^2 \overline{\langle \delta Q_{12} \delta Q_{34} \delta Q_{56} \rangle}\,,
\end{eqnarray*}
where overlap fluctuations can be written in terms of independent real replicas
with the same quenched disorder
\begin{equation}
\delta Q_{ab} \equiv \frac1N \sum_i s_i^a s_i^b- \frac1N \sum_i \overline{\langle s_i \rangle^2}\,.
\end{equation}
Each correlator $\W_i$ requires a number of different real replicas
equal to the largest index in its expression (right hand side).

Hence, we recall, we need two replicas to compute the overlap, four for the
susceptibilities and six for $\omega_1$ and $\omega_2$.

Can we use use a smaller number of replicas? The theory predicts
that there are six linear combination of the $\W_i$'s that diverge
less than the $\W_i$ separately.
Using these linear relationships one can express the eight
coefficients in terms of only the three-replicas 
estimators~\cite{parisi:13}:
\begin{eqnarray}
\omega_1^{(3)} & \equiv &
\frac{11}{30} \W_1- \frac{2}{15} \W_2\,,\\
\omega_2^{(3)} & \equiv &
\frac{4}{15} \W_1- \frac{1}{15} \W_2\,.
\end{eqnarray}
Alternatively the
theory predicts that there are three linear combination of the $\W$'s
that remain finite at the critical temperature. Therefore one can
express $\W_7$ and $\W_8$ as a function of the remaining cumulants
obtaining the four-replicas estimators \cite{veca:20}:
\begin{eqnarray}
\omega_1^{(4)}\equiv \frac{23 \W_1}{30}+\frac{\W_2}{20}-\frac{3
  {\W_3}}{5}+\frac{9 {\W_4}}{20}-\frac{6
  {\W_5}}{5}+\frac{{\W_6}}{2}\,, \nonumber \\
\omega_2^{(4)} \equiv
\frac{7 {\W_1}}{15}+\frac{2 {\W_2}}{5}-\frac{9 {\W_3}}{5}+\frac{3
  {\W_4}}{5}-\frac{3 {\W_5}}{5}+{\W_6}\,.  \nonumber \end{eqnarray}
Within the RS theory, the three- and four-replicas estimators are different from
the true $\omega_1$ and $\omega_2$ at any given temperature but {\it
  coincide} with them at the critical temperature. 
At a generic temperature $w_{1,r}$, $w_{2,r}$ and $\lambda$ have
model-dependent values and we are only interested in the universal
values they take at the critical temperature.  More precisely one can
show that close to the critical point
\begin{equation*}
\omega_i-\omega_i^{(3)} = O(|T-T_c|^{\gamma_{\Delta}})\ , \,\,\, \,  \omega_i-\omega_i^{(4)} = O(|T-T_c|^{\gamma_3}) \,,
\end{equation*}
where the exponent $\gamma_{\Delta}$ is expected to be smaller than
$\gamma_3$ (e.g.\ in MF one finds $\gamma_{\Delta}=1$ and $\gamma_{3}=3$).

\section{Finiteness of the longitudinal susceptibility.}
\label{appC}

Let us consider the model in presence of a Gaussian magnetic field
which generates a new term in the Hamiltonian: $+h_o \sum_i h_i S_i$,
where $h_i$ are independent Gaussian variables with zero mean and unit
variance.  The staggered magnetization is defined as
\begin{equation}
m_\mathrm{st}
\equiv \overline{\langle h_i \sigma_i \rangle}
\end{equation}
where $\overline{(\cdots)}$ is the joint average over the couplings
and the Gaussian magnetic field.  Its susceptibility is
\begin{equation}
  \chi_\mathrm{st}=\frac{\partial
  m_\mathrm{st}}{\partial h_0}=-\beta \sum_l \left(\overline{\langle
  h_i S_i h_l S_l\rangle - \langle h_i S_i \rangle \langle h_l S_l
  \rangle}\right)\,.
  \label{eq:stsus}
\end{equation}
Integrating by parts Eq. (\ref{eq:stsus})  one can finally obtain that
$\chi_\mathrm{st}=2 \beta \chi_L$.  Therefore, if the magnetic
susceptibility does not diverge, neither does the longitudinal
susceptibility.

\section{The models.}
\label{appD}

We study the $4D$-dimensional EA  model in a field $h$
where $N=L^4$ Ising spins interact via
\begin{equation}
  \label{eq:H}
{\cal H}= - \sum_{\langle xy\rangle} J_{xy} S_x\, S_y +
h \sum_x S_x \,,
\end{equation}
where the first sum is over nearest-neighbor pairs and $J_{xy} =\pm
1$ with 50\% probability. In our $4D$ computation, the spins
are located in the nodes of a hypercubic lattice with periodic
boundary conditions.

We have also simulated the model on a Bethe lattice where the spins
occupy the vertices of a random-regular graph with connectivity 4.

\bibliography{biblio}

\end{document}